\documentclass[twocolumn,prl,showpacs,letter]{revtex4}

\usepackage{graphicx}   
\usepackage{dcolumn}    
\usepackage{bm}         
\usepackage{color}

\begin{document}

\title{Observation of magneto-electric non-reciprocity in molecular nitrogen gas}

\author{B.~Pelle$^{1,2}$}

\author{H. Bitard$^{1,2}$}

\author{G. Bailly$^{1,2}$}

\author{C. Robilliard$^{1,2}$}

\email{cecile.robilliard@irsamc.ups-tlse.fr}

\affiliation{ $^{1}$Universit\'e de Toulouse; UPS; Laboratoire Collisions Agr\'egats R\'eactivit\'e, IRSAMC; F-31062 Toulouse, France.\\
$^2$CNRS; UMR 5589; F-31062 Toulouse, France.}

\date{\today}

\begin{abstract}
We report the direct observation of the non-reciprocity of the velocity of light, induced in a gas by electric and magnetic fields. This bilinear magneto-electro-optical phenomenon appears in the presence of crossed electric and magnetic fields perpendicular to the light wavevector, as a refractive index difference between two counterpropagating directions. Using a high finesse ring cavity, we have measured this magneto-electric non-reciprocity in molecular Nitrogen at ambient temperature and atmospheric pressure; for light polarized parallel to the magnetic field it is $2\eta_{\parallel \;{\mathrm{exp}}}(\mathrm{N_2})= (4.7\pm 1)\times 10^{-23}\;\mbox{m.V}^{-1}\mbox{.T}^{-1}$ for $\lambda =1064\;$nm, in agreement with the expected order of magnitude. Our measurement opens the way to a deeper insight into light-matter interaction, since bilinear magneto-electric effects correspond to a coupling beyond the electric dipole approximation. We were able to measure a magneto-electric non-reciprocity as small as $\Delta n = \left( 5\pm 2 \right) \times 10^{-18}$, which makes its observation in quantum vacuum a conceivable challenge.

\end{abstract}


\maketitle

Electro- and magneto-optical effects such as Faraday, Kerr, Pockels and Cotton-Mouton effects have been studied for more than a century, in gases as well as in condensed media. These studies resulted both in a better understanding of the interaction of light and matter, and in widely used applications. All these phenomena can be described in the framework of the electric dipole approximation of the light-matter interaction Hamiltonian, even if higher order terms may yield a significant contribution in some cases.

More recently, the interest in electro- and magneto-optical effects has been renewed by several experimental \cite{Roth2000,Roth2002,Rikken2002,Budker2003,Guena2005} and theoretical \cite{Baranova1977,Ross1989,Graham1983_84,Rizzo2003,Mironova2006,Chernushkin2008,DeLorenci2008,DeLorenci2010} studies concerning effects due to higher order terms of the light-matter interaction Hamiltonian, with particular emphasis on those bilinear in electric and magnetic fields. They are described, at the lowest order, by products between the electric dipole term on the one hand, and electric quadrupole or magnetic dipole terms on the other hand. However, their precise theoretical description is still somewhat controversial \cite{Baranova1977,Ross1989,Graham1983_84} and the predicted effect is 3 to $10^3$ times larger than observed \cite{Roth2000}: obviously, new measurements are needed.

From the experimental point of view, the corresponding effects are expected to be weaker than electric dipole ones by roughly a factor $\alpha$, $\alpha$ being the fine structure constant, i.e. two orders of magnitude. This makes them even more difficult to measure in dilute media than Kerr and Cotton-Mouton constants.

Recently, several of these effects have been observed in dense media \cite{Roth2000,Roth2002,Rikken2002}, either crystals or liquids, and several groups are working on potential applications in optics \cite{Kida2006,Saito2008,Igarashi2009}. Besides, the corresponding theoretical calculations are on their way \cite{Rizzo2009}. Measurements in gases are complementary to these studies in dense media: the interactions between atoms or molecules are most often negligible, and an ab-initio calculation of these properties is feasible with a good precision \cite{Rizzo2003}, thus enabling a detailed test of our understanding of these effects. In recent years, magneto-electric Jones dichroism has been observed in two atomic parity violation experiments \cite{Budker2003,Guena2005}, where it may generate systematics. Some calculations performed in atomic gases such as alkaline \cite{Chernushkin2008} and alkaline-earth-like atoms \cite{Mironova2006} have confirmed that bilinear magneto-electro-optical effects are far from negligible near some of the atomic resonances. Static electric and magnetic fields may thus become a novel tool for a fine control of the optical properties of the atomic media used in high precision measurements and metrology.

In this letter, we present a measurement of magneto-electric non-reciprocity in Nitrogen molecular gas. To our knowledge, it is the first time that such a bilinear magneto-electric dispersive effect is measured in dilute media, and it is also one of the smallest differences in refraction index ever measured ($\Delta n = \left( 5\pm 2 \right) \times 10^{-18}$), which demonstrates the potential of the ring cavity method, as suggested by \cite{Ross1989,Vallet2001}. 

In the presence of external crossed electric and magnetic fields, $\mathbf{E}$ and $\mathbf{B}$ respectively, the light velocity is no longer isotropic, whatever the propagation medium. More precisely, a light beam with wavevector $\mathbf{k}$, oriented along $\mathbf{B} \times \mathbf{E}$, experiences a refractive index that depends both on its polarization and its direction of propagation. This bilinear contribution can be written

\begin{equation}
n_i \left( \bm{\kappa}, {\mathbf{B}},{\mathbf{E}} \right) = \eta_i \;\, \left( {\mathbf{B}} \times {\mathbf{E}} \right) \cdot \bm{\kappa}
\end{equation}

\noindent where $i=\parallel,\perp$ refers to the angle between the light polarization and the magnetic field, and $\bm{\kappa} = \mathbf{k} / \left\| \mathbf{k} \right\|$ is the unitary vector parallel to $\mathbf{k}$. Therefore, two beams that counterpropagate in crossed and transverse E and B fields experience a refractive index difference $ \Delta n_i =2\eta_iEB$. This effect has been first observed on the imaginary part of the refractive index of a rare earth doped crystal \cite{Rikken2002}, where the relative anisotropic absorption was on the order of a few ppm on resonance with the Er$^{3+}$ ion, in an Er:YAG crystal. To our knowledge, up to now this effect has always been observed as a nonreciprocal dichroism \cite{Kida2006,Saito2008,Igarashi2009}, which is related to birefringence through Kramers-Kronig relations. Calculations have been made for atomic gases of alcaline-earth-like atoms near S-P or S-D resonances \cite{Mironova2006} and alcaline atoms near S-D resonances \cite{Chernushkin2008}. All these results are consistent with the expectation that for symmetry reasons, magneto-electric non-reciprocity (MENR) should have the same order of magnitude as magneto-electric Jones birefringence (MEJB), a birefringence with its eigenaxis oriented at $\pm 45^\circ$ with respect to the applied fields. In simple gases such as Nitrogen, Jones birefringence has been computed \cite{Rizzo2003} at a wavelength of $633\;$nm: after conversion for ambient temperature and atmospheric pressure, $\eta_{\mathrm{Jones\; theo}} (\mathrm{N_2}) = 9.0\times 10^{-23}\;$m.V$^{-1}$.T$^{-1}$.

\begin{figure}[htb]
\begin{center}
\includegraphics[width=8cm]{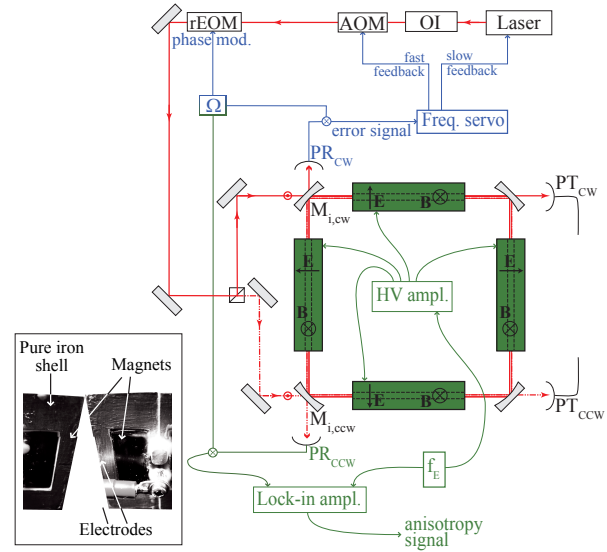}
\caption{(Color online) Scheme of our experimental set-up. The beampath is represented in red with arrows indicating the propagation direction, the frequency stabilization system in blue and the measurement signal generation in green. The blocks on each cavity arm represent the rods generating the magnetic and electric fields for the magneto-electric effects. The electric field is provided through a high-voltage amplifier (HV ampl) supplied with a sinusoidal wave at frequency $f_E$. Detection is made with a lock-in amplifier at frequency $f_E$. An optical isolator (OI) prevents feedback noise; the laser beam frequency is then frequency-shifted with an acousto-optic modulator (AOM). A resonant electro-optic modulator (rEOM) provides the phase modulation at angular frequency $\Omega$ for the Pound-Drever-Hall frequency stabilization. The servo actuators are the laser thermo-electric cooler and piezo-electric transducer as well as the AOM. Light is injected into the cavity both in the clockwise (cw) and counterclockwise (ccw) directions with a s-polarization; the $\mathrm{PR_{cw}}$ and $\mathrm{PR_{ccw}}$ (resp. $\mathrm{PT_{cw}}$ and $\mathrm{PT_{ccw}}$) photodiodes monitor the reflected (resp. transmitted) power in both directions. Inset: Picture of the rods generating $\bm{E}$ and $\bm{B}$.} \label{fig:Setup}
\end{center}
\end{figure}

In our experiment, we use a resonant passive ring cavity to convert $\Delta n_\parallel$ into a resonance frequency difference between counterpropagating beams. The set-up is somewhat similar to the one used in \cite{Hall2000}. Our apparatus has been described in details in \cite{Bailly2010,Robilliard2011}, and it is sketched in Figure \ref{fig:Setup}. It consists of a square resonant cavity with a perimeter $L=4L_0=1.6\;$m. The finesse varies between $15000$ and $50000$, depending on the quality of the alignment and the cleanliness of the mirrors, resulting in a cavity linewidth (FWHM) between 13 and 4~kHz. Light from a monolithic NonPlanar Ring Oscillator \cite{Kane1985} Nd:YAG laser (Wavelength Electronics NPRO 126, with a laser linewidth of 5~kHz in 1~ms) is injected in both clockwise and counterclockwise propagation directions. The laser frequency is stabilized on the clockwise resonance frequency thanks to the well-known Pound-Drever-Hall technique \cite{Drever1983}. In this technique, the light is phase-modulated at an angular frequency $\Omega$ much larger than the cavity linewidth, and an error signal proportional to the frequency detuning to the cavity resonance is extracted from the beam reflected from the cavity injection mirror $M_{i,\mathrm{cw}}$. We also extract a similar error signal in the counterclockwise direction. 

In our cavity, MENR superimposes to the Sagnac effect due to Earth rotation, which generates a resonance frequency difference $\Delta \nu_{S} = \frac{L_0}{\lambda}\Omega_{\mathrm{ER}} \cos \theta \simeq 20\,$Hz, where $\Omega_{\mathrm{ER}}$ is the Earth rotation angular velocity and $\theta \simeq 43^{\circ}$ the latitude of the laboratory. Therefore, the counterclockwise beam is slightly detuned from resonance, even in the absence of any magneto-electric effect. This detuning, which is small with respect to the cavity linewidth, appears as a DC non-zero counterclockwise error signal. Actually, the Sagnac effect is much larger than the expected MENR effect and might completely mask it, but we can distinguish the latter by modulating the electric field at frequency $f_E$ and detecting the $f_E$ frequency component in the error signal with a lock-in amplifier. 

The MENR effect is generated by 4 rods of length $L_{\bm{E}\times \bm{B}}=20\;$cm, which combine NdFeB permanent magnets providing a constant transverse magnetic field $B=0.85\;$T parallel to the light polarization and a pair of floating electrodes supplied by a high voltage amplifier (see inset of Fig. \ref{fig:Setup}): with a gap of $4\;$mm and a voltage up to $2\;$kV, the electric field can reach $0.5\;$MV/m. During an experimental run, one of the electrodes is grounded while the other one is connected to high voltage: the direction of the field can be inverted for each individual rod, simply by inverting the electrodes connection. More details on the rods are given in \cite{Robilliard2011}.

Calibration of the error signal in terms of frequency difference is made with an extra EOM placed on the clockwise beam just before the cavity mirror $M_{i,\mathrm{cw}}$ \cite{Bailly2010}. A sinusoidal voltage of frequency $f_E$ fed to this EOM results in a phase modulation, hence a frequency modulation of the sole clockwise beam. This mimicks the MENR effect and allows a precise calibration of the experiment. 

The frequency difference $\Delta \nu_{f_E}$ is finally related to $2\eta_\parallel$ by

\begin{equation}
2\eta_\parallel= \frac{\Delta \nu_{f_E}}{\nu} \frac{L}{L_{\bm{E}\times \bm{B}}} \frac{1}{\sum_{i=1}^4 \left( \bm{B}_i \times \bm{E}_i \right)\cdot\bm{\kappa}}
\label{eq:eta}
\end{equation}

\begin{figure}[htb]
\begin{center}
\includegraphics[width=8cm]{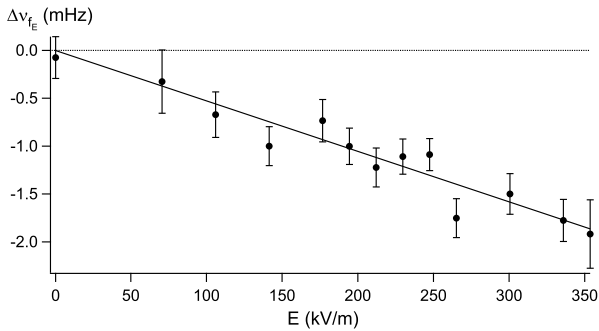}
\caption{Resonance frequency difference measured in N$_2$ as a function of the electric field, with 4 rods (2 with an upward B and the outer electrode grounded, and the other 2 with a downward B and the inner electrode grounded). The fitted slope is $(-5.27\pm 0.26)\times 10^{-9}\;$Hz.m.V$^{-1}$. The error bars represent one standard deviation and include only the statistical uncertainties. Each point corresponds to a 2000 s long run.} \label{fig:Eseries}
\end{center}
\end{figure}

As we have developed only a preliminary set-up, the ring cavity is contained in an almost air-tight plexiglas box, itself contained in a larger wooden box internally covered with Strasonic foam for thermal and noise insulation. Since we have no vacuum tank, the gas must be at atmospheric pressure: by using a continuous flow of N$_2$, we can work with almost pure N$_2$ in the laser cavity (estimated purity $\sim 98\; \%$). For many different connections of the rods, we have measured the frequency difference between the two counterpropagating directions as a function of the electric field $E$. Some typical results are presented in Fig. \ref{fig:Eseries}: the frequency difference is on the order of one mHz, with statistical uncertainties on the order of $200\;\mu$Hz for each point. It increases linearly with $E$, and it is proportional to the number of connected rods. Inverting the electric field leads to an opposite signal, as expected from an E-linear effect. 

Several tests have been made to check for fake effects. Firstly, the use of a lock-in amplifier at frequency $f_E$ eliminates most candidates: only E-linear effects might perturb our measurement. Secondly, we performed measurement series as a function of $E$ (see Fig. \ref{fig:Eseries}) in many different rods configurations: the results, a sample of which is presented in Table \ref{tab:tests}, were all consistent with the expected symmetries. The last two lines are of particular interest: indeed, they correspond to situations where the rods' effect cancels each other, so that the global result is null. 

\begin{table}[htb]
\begin{center}
\begin{tabular}{|c|c|c|c|c|}
\hline
N. rods & $E$ config. & B config. & Relative effect & Expected value \\ \hline
4 & $+\,+\,+\,+$ & $+\,+\,+\,+$ & 1 & 1 \\
4 & $-\,-\,-\,-$ & $+\,+\,+\,+$ & $-1.08\pm 0.21$ & -1\\
4 & $+\,+\,-\,-$ & $+\,+\,-\,-$ & $+0.92\pm 0.19$ & +1\\
3 & $+\,+\,-\,0$ & $+\,+\,-\,-$ & $+0.85\pm 0.24$ & +0.75\\
2 & $-\,-\,0\,0$ & $+\,+\,+\,+$ & $-0.50\pm 0.09$ & -0.5\\
2 & $0\,+\,0\,-$ & $+\,+\,-\,-$ & $+0.47\pm 0.11$ & +0.5\\
1 & $+\,0\,0\,0$ & $+\,+\,+\,+$ & $+0.27\pm 0.04$ & +0.25\\
2 & $+\,-\,0\,0$ & $+\,+\,+\,+$ & $+0.07\pm 0.2$ & 0\\
4 & $+\,+\,+\,+$ & $+\,+\,-\,-$ & $+0.16\pm 0.13$ & 0\\ \hline
\end{tabular}
\caption{Results of typical tests of the expected properties of MENR. The fields configuration of line 1 is the reference: all magnetic fields are oriented upward (+) and the grounded electrode for all rods is the inner one (+). The last two lines correspond to configurations where the effect of the rods cancels each other, resulting in a null global value.} \label{tab:tests}
\end{center}
\end{table}

\begin{figure}[htb]
\begin{center}
\includegraphics[width=8cm]{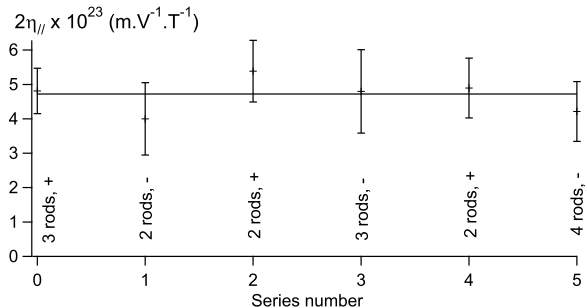}
\caption{Values of $2\eta_\parallel(N_2)$ for the different runs in Nitrogen made over a six week period. The error bars correspond to one standard deviation and include the statistical and calibration uncertainties.} \label{fig:eta}
\end{center}
\end{figure}

We present in Fig. \ref{fig:eta} the values of $2\eta_\parallel$ obtained in Nitrogen for different experimental runs made over a six week period. The error bars on the graph combine the statistical error and the $10\,\%$ uncertainty on the calibration factor. The weighted average of these values is $(4.7\pm 0.4)\times 10^{-23}\;m.$V$^{-1}$.T$^{-1}$. An extra uncertainty originates from the magnetic and electric fields determination, so that the final value in Nitrogen is $$2\eta_{\parallel\;\mathrm{exp}}(\mathrm{N_2})= (4.7\pm 1)\times 10^{-23}\;\mbox{m.V}^{-1}\mbox{.T}^{-1}.$$

As expected, this number is on the same order as the computed value $\eta_{\mathrm{Jones\; theo}} (\mathrm{N_2}) = 9.0\times 10^{-23}\;\mbox{m.V}^{-1}\mbox{.T}^{-1}$ obtained for Jones magneto-electric birefringence of Nitrogen at $\lambda =633\;$nm, $P=1\;$bar and $T=300\;$K after applying the appropriate pressure and temperature corrections for ideal gases. We summarize in Table \ref{tab:N2values} the experimental values of electro-and magneto-optical effects in N$_2$, along with those obtained from quantum chemistry calculations: they are in good agreement, but general arguments \cite{Baranova1977,Ross1989,Graham1983_84} suggest that MEJB and MENR constants should be on the order of $\alpha \sim 1/137$ times the square root of the product of Kerr and Cotton-Mouton constants, while they are substantially smaller ($\sim 30$), as was also the case in dense media \cite{Roth2000}.

\begin{table}[htb]
\begin{center}
\begin{tabular}{|c|c|c|}
\hline
Effect & Exp. value & Num. value \\ \hline\hline
Kerr (m$^2$.V$^{-2}$) & $1.4\times 10^{-25}$ \cite{Durand2009} & $1.6\times 10^{-25}$ \cite{Rizzo2005} \\\hline
Cotton-Mouton (T$^{-2}$) & $-2.1\times 10^{-13}$ \cite{Mei2009} & $-2.6\times 10^{-13}$ \cite{Cappelli2001} \\\hline
MEJB (m.V$^{-1}$.T$^{-1}$) & not available & $9.0\times 10^{-23}$ \cite{Rizzo2003} \\\hline
MENR$_\parallel$ (m.V$^{-1}$.T$^{-1}$) & $4.7\times 10^{-23}$ &  not available \\ \hline
\end{tabular}
\caption{Typical values of the main magneto- and electro-optical effects in N$_2$ at atmospheric pressure and ambient temperature. When available, we have considered the values at $\lambda = 1064\;$nm, as in our experiment, otherwise we have averaged the various published values.} \label{tab:N2values}
\end{center}
\end{table}

As a conclusion, we have measured magneto-electric non-reciprocity in Nitrogen molecular gas; this is to our knowledge the first observation of such a dispersive bilinear magneto-electric effect in a gas, far from any resonance line. The next step will be to use rotatable rods to measure MENR for different fields orientations. Applying parallel electric and magnetic fields will also allow us to measure magneto-electric Jones birefringence and check that it is indeed equal to $\eta_\perp + \eta_\parallel$, as it should be \cite{Graham1983_84,Robilliard2011}. A vacuum tank presently under construction will allow us to improve our measurement precision and to study different gases, among which atomic gases such as Krypton and Xenon, where relativistic effects are expected to be significant. Our experiment demonstrates that many effects that were previously beyond experimental reach can now be measured in quite reasonable timescales. Our long term goal is to search for the magneto-electric non-reciprocity of quantum vacuum \cite{Rikken2003, Robilliard2011}, which is approximately $7\times 10^8$ times smaller than what we have measured. Its detection would require fields as high as $B=15\;$T and $E=20\;$MV/m, a better cavity with a finesse of $200\,000$ and a noise level corresponding to the shot-noise level with an injected laser power near 50 mW. All these performances have already been achieved separately, but bringing them together is obviously a very ambitious challenge.

The authors are grateful to Jacques Vigu\'e for discussions and support, and to A. Aspect, F. Bretenaker and B. Girard for useful suggestions on the manuscript. They also thank the technical staff from LCAR for their helpfulness.


\end{document}